\documentclass[10pt,letterpaper]{article}
\usepackage{opex3}

\usepackage{graphicx}
\usepackage{amssymb}
\usepackage{amsmath}
\usepackage{mathrsfs}
\newcommand{\ud}{\mathrm{d}}

\newcommand{\ket}[1]{|#1\rangle}

\newcommand{\de}{\partial}

\newcommand{\eps}{\epsilon}

\usepackage[ps2pdf, breaklinks=true]{hyperref}

\begin{document}
%% NOTE: TITLE PAGE & TOC NOT USED FOR MANUSCRIPT SUBMISSIONS %%
\title{Tunable Modulational Instability Sidebands via Parametric Resonance in Periodically Tapered Optical Fibers}
\author{Andrea Armaroli and Fabio Biancalana}
\address{Max Planck Research Group 'Nonlinear Photonics Nanostructures' \\ Max Planck Institute for the Science of Light,
G{\"u}nther-Scharowsky-Str.~1/Bau 24\\
91058 Erlangen, Germany}
\email{andrea.armaroli@mpl.mpg.de} %% email address is required
%\homepage{http://mpl.mpg.de/mpf/php/abteilung3/jrg/home/} %% author's URL, if desired

%%%%%%%%%%%%%%%%%%% abstract and OCIS codes %%%%%%%%%%%%%%%%
%% [use \begin{abstract*}...\end{abstract*} if exempt from copyright]

\begin{abstract} \!\!\!\!\!\!\!\!\! We analyze the modulation instability induced by periodic variations of group velocity dispersion and nonlinearity in optical fibers, which may be interpreted as an analogue of the well-known parametric resonance in mechanics. We derive accurate analytical estimates of resonant detuning, maximum gain and instability margins, significantly improving on previous literature on the subject. We also design a periodically tapered photonic crystal fiber, in order to achieve narrow instability sidebands at a detuning of $35$ THz, above the Raman maximum gain peak of fused silica. The wide tunability of the resonant peaks by variations of the tapering period and depth will allow to implement sources of correlated photon pairs which are far-detuned from the input pump wavelength, with important applications in quantum optics.
\end{abstract}


%%%%%%%%%%%%%%%%%%%%%%% References %%%%%%%%%%%%%%%%%%%%%%%%%
%\bibliographystyle{osajnl}
%\bibliography{../ParametricResonance3}

%%%%%%%%%%%%%%%%%%%%%%%%%%%%%%%%%%%%%%%%%%%%%%%%%%%%%%%%%%%

\section{Introduction and motivations}

Parametric resonance (PR) is a well-known instability phenomenon which occurs in systems the parameters of which are varied periodically during evolution, see the classical treatments in Refs. \cite{ArnoldCM,LandauCM}. For example, a harmonic oscillator the frequency of which is forced to vary in time, will become unstable if its internal parameters and the amplitude of the frequency variation happen to be inside special regions, known as {\em resonance tongues}. The study of the properties of resonance tongues has a long history and relies on a variety of geometrical approaches \cite{ArnoldGMODE,Broer2000}.

It is natural that such a general phenomenon was associated to the equally important instability process that is ubiquitous in infinite dimensional dynamical systems: modulation instability (MI), also known as Benjamin-Feir instability \cite{BenjaminFeir,BespalovTalanov}. MI is known to exist in different branches of physics such as fluid-dynamics \cite{Whitham1965}, plasma physics \cite{Taniuti1968,Hasegawa1970,Tam1969}, Bose-Einstein condensates \cite{Konotop2001} and solid-state physics \cite{Sievers1998}. In nonlinear optics \cite{Karpman1967}, it manifests itself as an exponential growth of two spectrally symmetric sidebands on top of a plane wave initial condition, by virtue of the interplay between the cubic Kerr nonlinearity and the group velocity dispersion (GVD). In optical fibers it leads to the breakup of a plane wave into a train of normal modes of the system, i.e.~solitons \cite{Hasegawa1986,Akhmediev1986}.

The link between PR and MI has been established in 1993, see \cite{Matera1993}, in relation to the periodic re-amplification of signals in long-haul telecommunication optical fiber cables. This was based on a nonlinear Schr{\"o}dinger equation (NLS) where the coefficient of the nonlinear term is varied along the propagation direction. Importantly, this peculiar type of MI occurs in both normal and anomalous GVD. This prediction was later partially verified in experiments, see \cite{Kikuchi1995,Tang2000}.

Moreover, in long-haul fibers, dispersion management is a commonly used technique which introduces periodic modulation of fiber characteristics. The possibility of instability phenomena disrupting adjacent communication channels has been thoroughly analyzed, see e.g.~\cite{Smith1996,Bronski1996,TchofoDinda2003,TchofoDinda2008}. Specifically, in \cite{Smith1996} the partial suppression of the conventional MI in anomalous GVD
due to a large swing dispersion management is discussed, while in \cite{Bronski1996} the degenerate case of zero average dispersion was studied. The combination of both loss and dispersion compensation is studied in \cite{TchofoDinda2003,TchofoDinda2008}. The main interest in those works was on  step-like variations of the GVD coefficient.

At the same time the effects of smooth periodic or random variations of fiber parameters were studied in \cite{Abdullaev1996,Abdullaev1997,Abdullaev1999}. 
Also some work has been done on the effect of the perturbation of fiber parameters on soliton propagation \cite{Bauer1995,Pelinovsky2004}.

It turns out that the variation of dispersion and nonlinearity can enhance or suppress the PR, while higher order nonlinear effects such as self-steepening proved less important. Quite surprisingly,  experiments on  micro-structured fibers have been reported only recently for the first time, see Refs.~\cite{Mussot2012,Mussot2012b}, where a  photonic-crystal fiber (PCF, \cite{RussellScience2003}) of varying diameter is used. In that experiment, the dispersion is periodically switched from normal to anomalous, but this feature is not required to achieve PR, while the effect of Raman scattering plays an important role in the relative magnitude of the PR peaks.

The conventional  explanation is in term of a grating-assisted phase matching process \cite{Matera1993, TchofoDinda2003, TchofoDinda2008, Mussot2012}, with the exception of Ref.~\cite{Abdullaev1997}, which relies on the theory of Mathieu equation ---a paradigmatic example of the application of Floquet theory, see Refs. \cite{LandauCM,ArnoldGMODE}---and analyzes the case of nonlinearity variations only.

In this work we present an improved analytical treatment based on the so-called Poincar{\'e}-Lindstedt perturbation  method and averaging method, which provide excellent predictions on resonant frequency, gain and bandwidth of the unstable peaks appearing in the output spectra. The new feature here is the tunability of the instability bands, specifically the lowest order one, which possesses a gain larger than the other PR peaks. Finally we provide physical estimates for a PCF-based system which leads to instability bands with frequency detunings larger than $13$ THz, the detuning of the Raman gain peak of fused silica. This system can be used for the generation of entangled photon pairs in a narrow bandwidth, far detuned from the Raman gain peak, which would reduce the impact of Raman-induced decorrelations, with important applications in quantum optics, especially in quantum computation and cryptography \cite{Rarity2005,Silberhorn2010}.

\section{Model equation and Floquet theory: analytical and numerical analysis}

\subsection{NLS with varying parameters and derivation of Hill equation}

Let us consider the dimensionless NLS equation for a slowly-varying electric field envelope
$A(z,t)$ of a guided mode at carrier frequency $\omega_0$, with both GVD and nonlinearity varying along the propagation direction, namely
\begin{equation} \label{NLS1}
i\de_{z}A- \frac{1}{2}s(z)\de_{t}^{2}A+n(z)|A|^2A=0.
\end{equation}
$s$ and $n$ are normalized coefficients, $s(z)\equiv\beta_2(z)/\beta_2^0$ and $n(z) \equiv \gamma(z)/\gamma^0$, where $\beta_2(z)$ and $\gamma(z)$ are the GVD and nonlinear coefficients, respectively, and the $0$ superscript denotes their mean values. $n$ and $s$  are assumed to be periodic functions of $z$. 
Finally $z\equiv Z/Z_{nl}$ is the dimensionless distance in units of the nonlinear length $Z_{nl}\equiv(\gamma^0 P_{t})^{-1}$, and $t\equiv(T-v_g^0 Z)/T_{s}$ is the dimensionless retarded time in units of  $T_{s}\equiv \sqrt{Z_{nl} |\beta_2^0|}$, and $v_g^0$ is the mean group velocity. $P_t$ is the total input power injected in the mode, and $A$ is the dimensionless modal intensity scaled by $\sqrt{P_{t}}$.

%($A$ is normalized so that it carries unit power).

We look for for a steady state solution of \eqref{NLS1} in the form $A=\sqrt{P_0}\exp{(i \phi(z))}$: it can be verified that $\phi(z) = P_0\int_{-\infty}^{z}{n(z')\ud z'}$. We then perturb this steady state by adding a small complex time dependent contribution $a(z,t)$, i.e.~$A(z,t)=\left(\sqrt{P_0}+\varepsilon a(z,t)\right)\exp{(i \phi(z))}$, with $\eps\ll 1$. Inserting this ansatz in Eq.~\eqref{NLS1} and taking only the terms which are first order in $\eps$, one finds that  $a$ obeys the following equation:
\begin{equation}
	i\partial_z a -\frac{1}{2}s(z)\partial^{2}_{t}a+n(z)P_0(a+a^*) = 0.
\label{eq:NLS1lin}
\end{equation}
Finally we substitute in \eqref{eq:NLS1lin} the {\rm ansatz}
\[
a(z,t) = a_A(z) e^{-i \delta t} + a_S(z) e^{i \delta t}
\]
to obtain the following Schr\"odinger equation
\begin{equation}
	i\dot{\ket{\psi}} = H_{\rm s}(z) \ket{\psi},\,\,\,
	H_{\rm s}(z) \equiv \left(-s(z)\frac{\delta^2}{2}-n(z) P_0\right)\hat{\sigma}_z - i n(z)P_0 \hat{\sigma}_y
%	=\begin{pmatrix}
%		-s(z)\frac{\delta^2}{2}-n(z)P_0 & -n(z)P_0\\
%		n(z)P_0 &	 s(z)\frac{\delta^2}{2}+n(z)P_0\\
%	\end{pmatrix}
	\label{eq:NLS1SAS}
\end{equation}
where the dot denotes $z$-derivative, $H_{\rm s}(z)$ is a non-Hermitian Hamiltonian (which thus allow unstable modes to grow), $ \ket{\psi} \equiv (a_A,a_S^*)^T$, and
 $\hat{\sigma}_i$ are the Pauli matrices.

In the remaining part of this section we will assume that dispersion and nonlinearity exhibit the simplest possible periodic behavior
\begin{equation}
	s(z) = s_0+\tilde s(z) = s_0+h s_1\cos{\alpha z}, \;	n(z) = n_0+\tilde n(z) = n_0+h n_1\cos{\alpha z},
	\label{eq:dispnlcos}
\end{equation}
where generally $s_0=\pm1$ for normal (anomalous) dispersion and $n_0=1$; $\alpha$ is the normalized spatial angular frequency for the parameter oscillation. The forcing amplitude is controlled by parameter $h$, which must be small to guarantee the validity of our perturbative expansions. However, we find below that our estimates are reliable even for $h\sim 0.5$.

The conventional way in which PR is approached in fiber optics is to split the dispersion coefficient into two parts: a constant average term and an oscillating term. By means of a change of variables this latter is included in a single varying nonlinear coefficient, see Refs.~\cite{Smith1996,TchofoDinda2003}. Then all quantities are expanded in Fourier series  and it is assumed that only one Fourier coefficient resonates at each PR order. This leads to the nonlinear phase-matching condition for the $m$-th peak resonant frequency
\begin{equation}
	s_0\delta_m^2 + 2 n_0 P_0 = m\alpha
\label{eq:grating}
\end{equation}
where $m$ is the PR order and $\delta_m$ the corresponding resonance detuning. This condition is valid for $m>0$ ($m<0$) if GVD is normal (anomalous). Below, in our derivation, we  assume instead that $m$ is a positive integer.
Eq.~\eqref{eq:grating} can also be understood as a grating assisted phase matching condition, the grating being the periodic modulation of the fiber parameters.
We performed detailed numerical simulations and verified that the result of \eqref{eq:grating} is quite coarse. The purpose of the present section is to prove that a more accurate estimate can be made, in a more general case than the variation of nonlinear coefficient only, which was already discussed in \cite{Abdullaev1997}.
We define the parameter space as the plane $(\delta,h)$ and we investigate the different features of PR instability (resonant frequency, bandwidth and gain) on this plane, for a fixed value of $|n_1|/|s_1|$.

In order to compare to the theory of PR in a classical mechanical oscillators, it is instructive to derive a 2nd order ODE from the system of Eqs.~\eqref{eq:NLS1SAS}. Let us first transform it in phase-quadrature variables, by applying the rotation
\[
R = \frac{1}{\sqrt{2}}
\begin{pmatrix}
1&1\\1&-1
\end{pmatrix}
\]
We define the new state $\ket{\phi}=R\ket{\psi}$ which satisfies a system analogous to that in Eq.~\eqref{eq:NLS1SAS}, with Hamiltonian matrix
\begin{equation}
	H_{\rm pq} = RH_{\rm s}R^{-1} =
	\begin{pmatrix}
0&c_1(z)\\c_2(z)&0
\end{pmatrix}
\label{eq:NLS1PQ}
\end{equation}
with $c_1(z) = -\frac{1}{2}s(z)\delta^2$ and $c_2(z) = c_1(z)-2n(z)P_0$. We can thus more easily derive a second order ODE for both $\phi_{1,2}$:
\begin{equation}
	\ddot{\phi}_{1,2}-\frac{\dot{c}_{1,2}}{c_{1,2}}\dot{\phi}_{1,2}+c_1c_2\phi_{1,2} = 0,
	\label{eq:linODE2}
\end{equation}
which still contains a first derivative of $\phi_{1,2}$. This can be eliminated by the substitution
\[
\phi_{1,2} = \exp{\left(\frac{1}{2}\int_0^z{\frac{\dot{c}(z')_{1,2}}{c(z')_{1,2}}\ud z'}\right)}\tilde{\phi}_{1,2} = \sqrt{c_{1,2}}\tilde{\phi}_{1,2}.
\]
We finally obtain the following ODE
\begin{equation}
	\ddot{\tilde{\phi}}_{1,2}
	+ \left\{c_1c_2 + \frac{1}{2}\frac{\ddot{c}_{1,2}}{c_{1,2}}
	- \frac{3}{4}\left[\frac{\dot{c}_{1,2}}{c_{1,2}}\right]^2\right\}
	\tilde{\phi}_{1,2}=0
\label{eq:NLSHill}
\end{equation}
Notice that in contrast to a usual dissipative term, which corresponds to a threshold in the magnitude of the perturbation needed to achieve PR instability (see Ref. \cite{LandauCM}), our variable transformation is simply an oscillating factor, which does not affect the instability.

The  left-hand side of \eqref{eq:NLSHill} contains, even in our simple case of Eq.~\eqref{eq:dispnlcos}, several harmonic terms. This means that Eq.~\eqref{eq:NLSHill} is a Hill equation, which generalizes the Mathieu equation found in \cite{Abdullaev1997}, where $\dot{c}_1=0$.  It is well known that Mathieu equation is an exceptionally simple case, \cite{ArnoldCM,Broer2000}, with a predictable structure of stability and instability regions in the parameter space.
Moreover since we consider a Hill equation, we expect the instability regions to be irregular and appear in the form of instability islands separated by stable parts. This is consistent with the statements made in Ref.~\cite{Smith1996}, i.e.~that a large switching of dispersion  may suppress instability.

Despite Eq.~\eqref{eq:NLSHill} highlights all these important properties of the system, it is difficult to handle analytically. We thus turn back to  \eqref{eq:NLS1PQ} and derive our analytical estimates from it, in order to report the simplest possible treatment.

In the following part we present the main results of our calculations, which rely on the commonly used methods of averaging and on the Poincar{\'e}–-Lindstedt perturbation method, see \cite{VerhulstEnc}.

\subsection{Estimate of Resonant Frequency and Parametric Gain}

We now explore the properties of the system
\begin{equation}
i\dot{\ket{\phi}} = H_{\rm pq}(z) \ket{\phi}
\label{eq:NLS1PQ2}
\end{equation}
with $H_{\rm pq}$ defined in \eqref{eq:NLS1PQ}. In the limit of vanishing perturbation we have a simple harmonic oscillator written in complex variables $\phi_{1,2}$. It is widely known that, in this limit, parametric resonance occurs if the natural frequency of the oscillator is a multiple of half the forcing frequency and this is the basic point behind any perturbative approach, see \cite{LandauCM}. Thus, by setting $c_{1,2}(z)=c_{1,2}^0+\tilde{c}_{1,2}(z)$,  the natural frequency of the unforced oscillator is simply $\omega_0=\sqrt{c_1^0c_2^0}$. The resonance condition becomes $\omega_0=m\alpha/2$, with $m=1,\,2,\,3,\ldots$ which expressed in terms of the NLS parameters corresponds to a detuning
\begin{equation}
	\delta_m = \frac{1}{\left|s_0\right|}\sqrt{2n_0P_0\left[-s_0 + \left|s_0\right|\sqrt{1+\left(\frac{m\alpha}{2n_0P_0}\right)^2}\right]}
	\label{eq:NLSreson1}
\end{equation}
Note that, if we assume $\left|\frac{m\alpha}{2n_0P_0}\right|\gg1$, we obtain the quasi-phase matching condition given in Eq.~\eqref{eq:grating} (taking care of the different convention on $m$, which in our relation, Eq.~\ref{eq:NLSreson1} is only positive, while in Eq.~\eqref{eq:grating} can also be negative), which is thus a coarse approximation if $\alpha\approx n_0P_0$.

In order to obtain the peak gain at resonance we use the method of averaging, which consists in posing 
\begin{equation}
\phi_1 = A(z) \cos{\omega_0 z} + B(z) \sin{\omega_0 z},\;
\phi_2 = -\frac{i\omega_0}{c^0_1}\left[ A(z) \sin{\omega_0 z} - B(z) \cos{\omega_0 z}\right],
\label{eq:avansatz}
\end{equation}
 substituting in \eqref{eq:NLS1PQ2} and averaging over a period of the forcing term $T_z \equiv 2\pi/\alpha$, we obtain
\begin{equation}
	\begin{pmatrix}
		\dot{A}\\\dot{B}
	\end{pmatrix}
	= \frac{\alpha  (\rho_2-\rho_1) \omega_0^2}{2 \pi  \left(\alpha ^2-4 \omega_0^2\right)}
	\begin{pmatrix}
	 -1+\cos\left(\frac{4 \pi  \omega_0}{\alpha }\right) &  \sin\left(\frac{4 \pi  \omega_0}{\alpha }\right)\\
 	  \sin\left(\frac{4 \pi  \omega_0}{\alpha }\right)& 1-\cos\left(\frac{4 \pi  \omega_0}{\alpha }\right)
	\end{pmatrix}
		\begin{pmatrix}
		A\\B
	\end{pmatrix}
	\label{eq:ave1}
\end{equation}
where $\rho_1 \equiv h s_1/s_0$ and $\rho_2 \equiv h(\delta^2 s_1+4n_1 P_0)/(\delta^2 s_0 +4 n_0P_0)$. It is thus  apparent that the $m=1$ resonance occurs at $2\omega_0=\alpha$ and in this case the matrix of the system \eqref{eq:ave1} has purely real eigenvalues of opposite sign. The positive one  corresponds to the peak gain of the first PR band and turns out to be the maximum achievable value. This reads as
\begin{equation}
	g_1 = \frac{\delta_1^2P_0}{2\alpha}h\left|s_0n_1-s_1n_0\right|
\label{eq:gain1}
\end{equation}
where $\delta_1$ is calculated according to Eq.~\eqref{eq:NLSreson1}.

We notice that there may exist a set of parameter values which suppresses or even forbids the occurrence of the instability, specifically $s_1/s_0=n_1/n_0$. This implies that, e.g., in normal GVD ($s_0=1$) if both nonlinearity and dispersion undergo parallel increase and decrease, the instability is suppressed. Instead the same condition maximizes the gain under anomalous GVD ($s_0=-1$).

The higher order resonances are more difficult to characterize by this method, but their gain is generally of order $h^2$, since in this case the first-order contribution vanishes, see Eq.~\eqref{eq:ave1}.

\subsection{Estimate of resonance bandwidth}

Floquet theory predicts that our system \eqref{eq:NLS1PQ2} has quasi-periodic solutions (composed by a periodic function with period $T_z$ and a phase-factor, a complex function of unit absolute value). Moreover, stability margins correspond to a pair of periodic or anti-periodic solutions, defined by $\ket{\phi(z+T_z)}=\pm\ket{\phi(z)}$, which possess periods equal to $T_z$ and $2T_z$ respectively. Bearing this in mind, we apply the Poincar{\'e}-Lindstedt method, by making the following perturbation expansion in powers of $h$:
\begin{equation}
 -\frac{1}{2}\delta^2 = d_0 + h d_1 + h^2 d_2,\;\ket{\phi(z)} = \ket{\phi_0(z)}+h \ket{\phi_1(z)}+h^2 \ket{\phi_2(z)}+\ldots, \;
 \ket{\phi_i}=(\phi_{1i},\phi_{2i})^T.
\label{eq:PLpert}
\end{equation}
We choose  $d_0$ such that it corresponds to the first resonant frequency, i.e.~$d_0=-\delta_1^2/2$.

At zeroth order we obtain
\begin{equation}
\ddot{\phi}_{10}+\omega_0^2\phi_{10}  = 0
\label{eq:PL0}
\end{equation}
%which has two linearly independent solutions $\phi_{10}=\cos{\omega_0z}$ and $\phi_{10}=\sin{\omega_0z}$. 
The first resonant band occurs at $\omega_0=\alpha/2$, thus Eq.~\eqref{eq:PL0} has solutions with period $2T_z$. For a fixed value of $h$, the stability margins correspond in general to different frequencies, which implies they must correspond to linearly independent eigenfunctions. The higher order corrections, which provide the stability margins, are obtained by solving for the successive terms of the perturbation series ($\ket{\phi_1(z)},\,\ket{\phi_2(z)},\ldots$) by making use alternatively each of the independent solutions of Eq.~\eqref{eq:PL0}.

We thus first pose $\phi_{10}=\cos{\omega_0z}$, and at first order in $h$ we obtain:
\begin{multline*}
\ddot{\phi}_{11}+\omega_0^2\phi_{11}  = \left[-d_1s_0\left(c_1^0+c_2^0\right) + \left(c_2^0-c_1^0\right)\frac{d_0s_1}{2}+c_1^0n_1 P_0\right]\cos{\omega_0z}+\\
\left[-\left(3c_2^0+c_1^0\right)\frac{d_0s_1}{2}+c_1^0n_1 P_0\right]\cos{3\omega_0z}.
\end{multline*}

We then impose that the secular term vanishes and solve for $d_1$,
\begin{equation}
d_1 = \frac{1}{2s_0\left(d_0s_0-n_0P_0\right)}d_0P_0(s_0n_1-n_0s_1).
\label{eq:bandwidth1}
\end{equation}

If we set $\phi_{10}=\sin{\omega_0z}$ we obtain the same value with opposite sign, so that the instability margins of the first band satisfy
\begin{equation}
\frac{\delta^2}{2}=  \frac{\delta_1^2}{2} \mp h d_1
\label{eq:bandedges}
\end{equation}
The four solutions of Eq.~\eqref{eq:bandedges} are denoted by $\pm\delta_1^{\pm}$, and the bandwidth (i.e. the range of unstable detuning values) is given by $\Gamma_1\equiv\delta_1^{+}-\delta_1^{-}$. 
Higher order corrections and bandwidth of overtone resonances can be found by solving for the higher order terms $\ket{\phi_1(z)},\,\ket{\phi_2(z)},\ldots$ of the perturbative expansion, but the calculation is too lengthy to report here.

\subsection{Comparison of analytical and numerical results of the Hill equation}

The Hill equation \eqref{eq:NLSHill} can be solved numerically by means of an ODE solver, following the prescriptions of the Floquet theory. The  monodromy matrix is calculated by setting two linearly independent initial conditions and evaluating the solution at $z=T_z$, see \cite{ArnoldGMODE}. The eigenvalues of the monodromy matrix provide the instability gain (Floquet exponents).

Since PR bands are typically narrow, this requires a fine grid in the $(\delta,h)$ parameter space. In order to speed-up the calculation we (i) compute the exact instability margins, then (ii) calculate the gain values of each instability tongue.

The calculation of the exact stability margins can be carried out by the standard Hill determinant method or harmonic balance based on the truncated Fourier expansion of the variables ($\phi_{1,2}$) and forcing terms, see \cite{Deconinck2006}. In contrast with more conventional spectral problems, where the eigenvalue appears  explicitly in the equation, we need to find the values of detuning $\delta$ at a fixed $h$, while the coefficients depend on $\delta$ in a nontrivial way.
This implicit dependence is solved by means of a root finding algorithm based on the minimization of the least singular value, see \cite{Labay1992a}.
The second step involve the numerical evaluation of the Floquet exponents in the close proximity of each band. We compare the results of the numerical calculation of resonant tongues and analytical estimates for the case of maximal gain and bandwidth, i.e.~$n_1/n_0=-s_1/s_0$, see \eqref{eq:gain1} and \eqref{eq:bandwidth1}.

\begin{figure}
	\centering
		\includegraphics[width=0.49\textwidth]{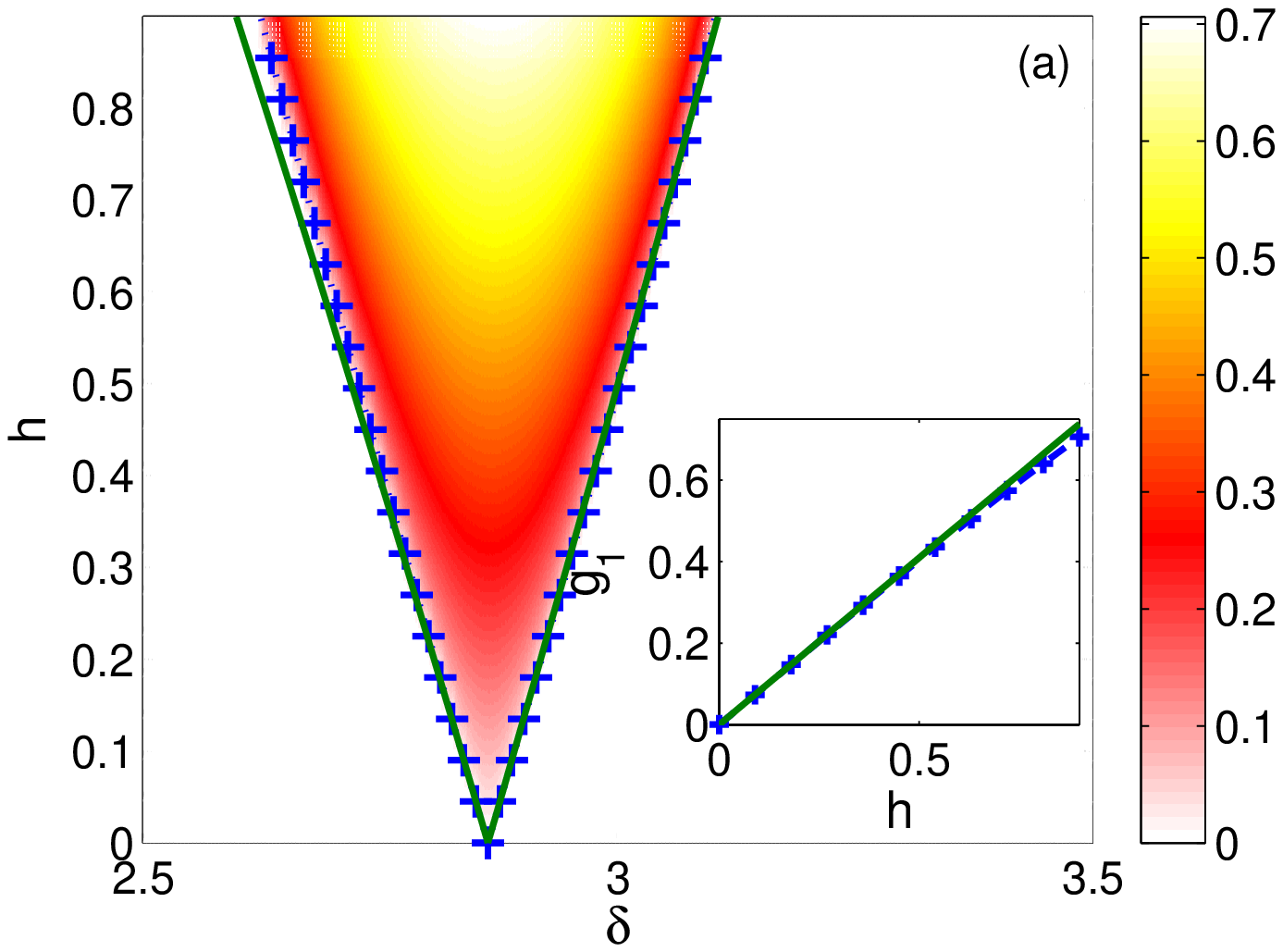}
		\includegraphics[width=0.49\textwidth]{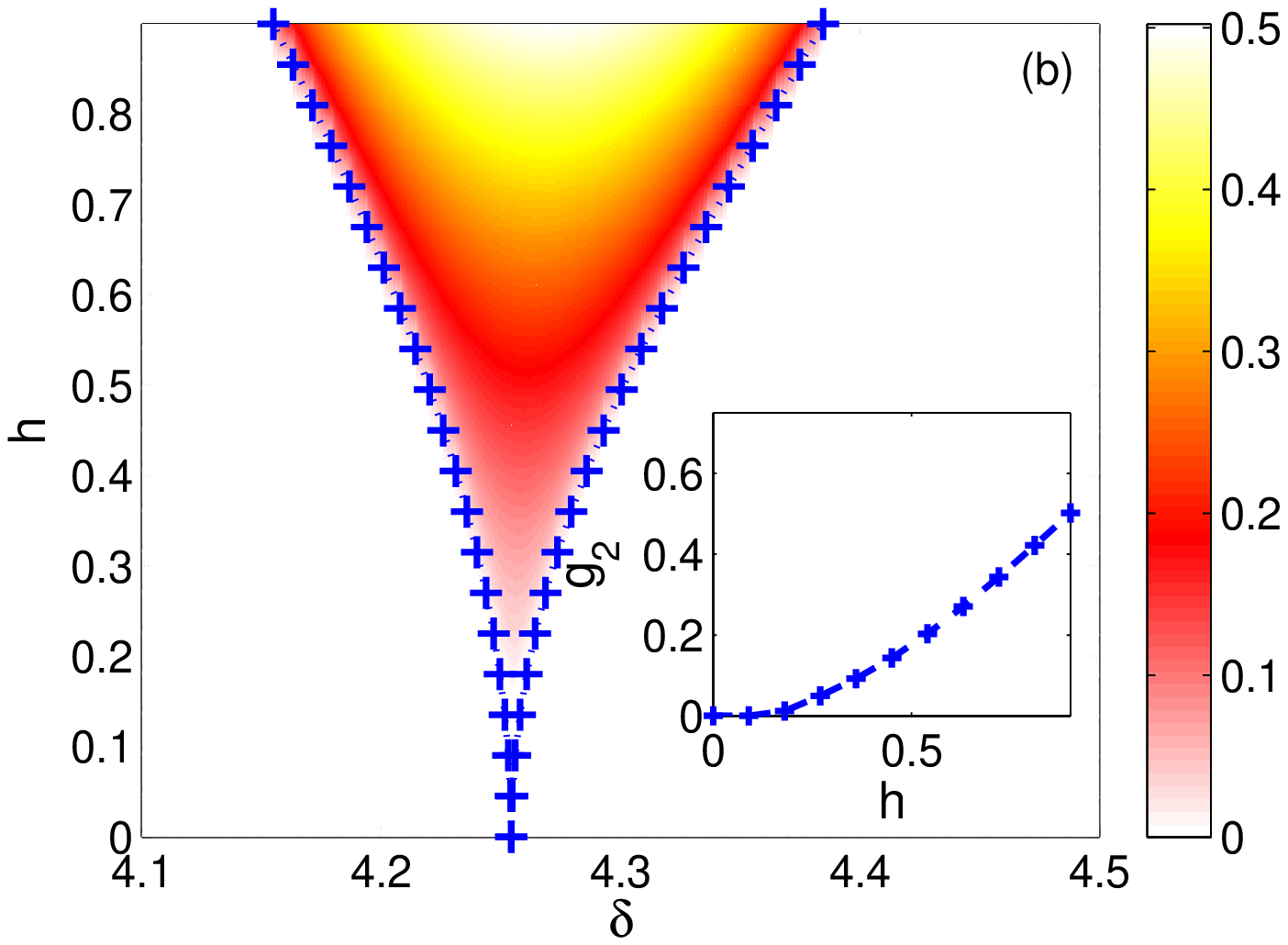}
\caption{Resonant tongues in the $(\delta,h)$ plane for the parametric resonance instability of a NLS with varying dispersion and nonlinearity, for (a) the first and (b) the second resonant bands. GVD is normal ($s_{0}=+1$) and $\alpha=10$, $n_1=-s_1=1$ are chosen in order to obtain the maximum gain and bandwidthm, see Eqs.~\eqref{eq:gain1} and \eqref{eq:bandwidth1}.  The colormap provides the values of gain computed by means of an ODE solver. The solid green lines are the analytic predictions of Eq.~\eqref{eq:bandedges} and the blue dotted lines with markers are the band edges computed numerically by employing the Hill determinant method (see in the text). The insets show the maximum gain vs.~$h$ (blue dotted line with markers) and in (a) also the  curve obtained from Eq.~\eqref{eq:gain1} (green solid). Notice that since we reported in the text only the perturbation results of the first PR tongue, in $(b)$ we show only numerical results.}
	\label{fig:alpha10normal}
\end{figure}

\begin{figure}
	\centering
		\includegraphics[width=0.49\textwidth]{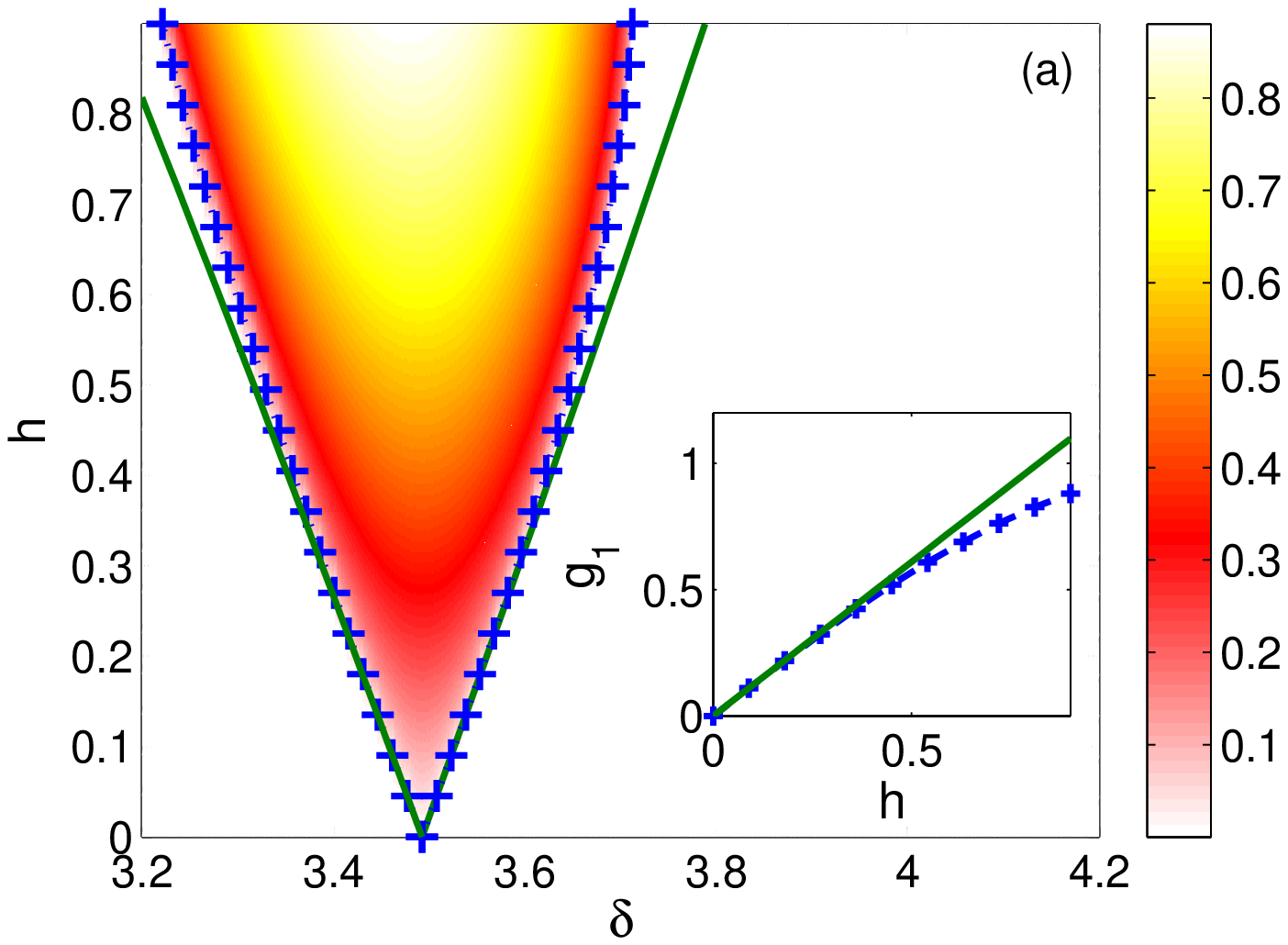}
		\includegraphics[width=0.49\textwidth]{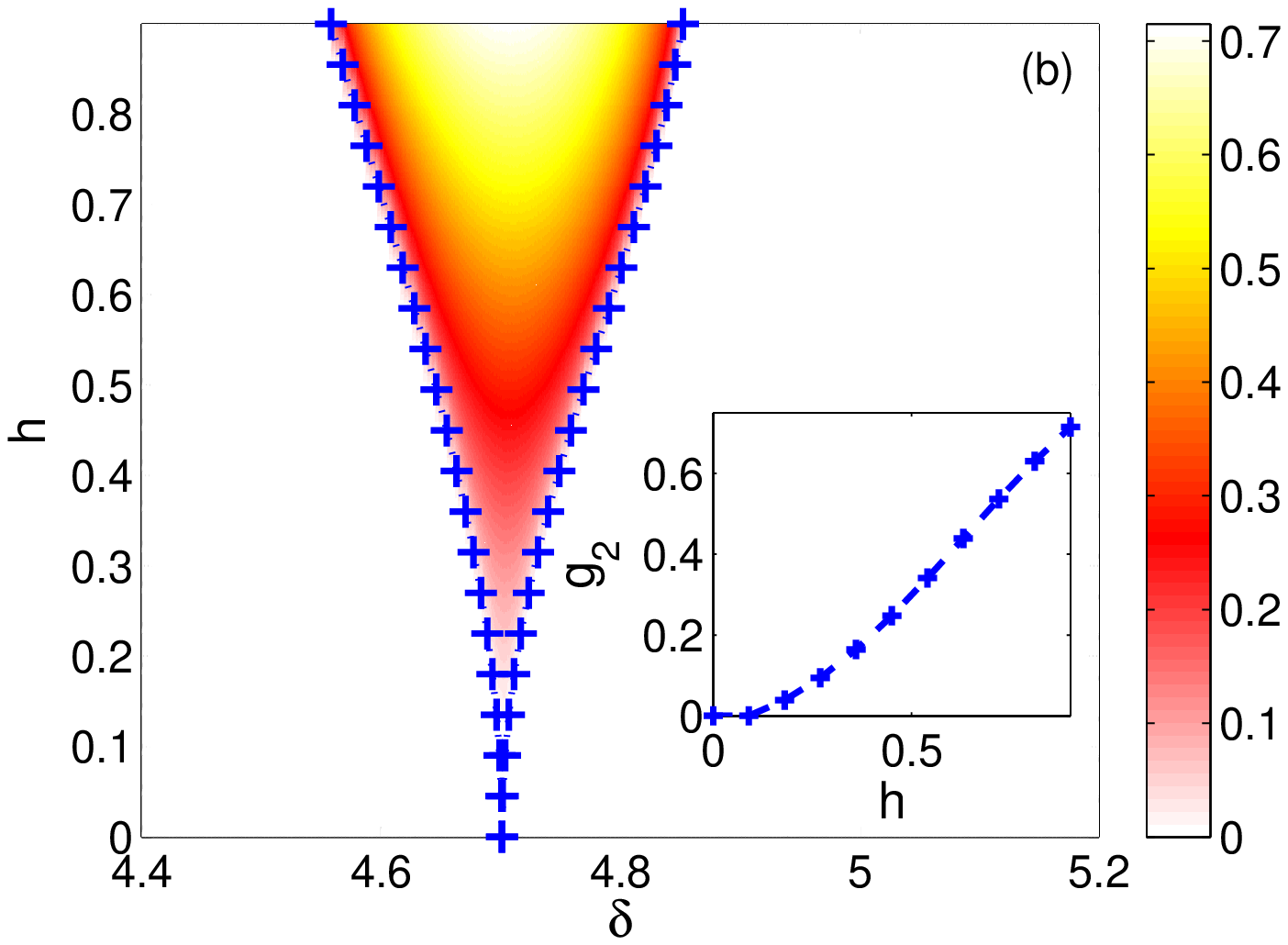}
	\caption{Same as in Fig.~\ref{fig:alpha10normal}, but in the anomalous GVD regime ($s_0 = -1$). In this  case $n_1=+s_1=1$.}
	\label{fig:alpha10anomal}
\end{figure}

First we show in Figs.~\ref{fig:alpha10normal} and \ref{fig:alpha10anomal} the instability regions for both normal and anomalous GVD at a fixed frequency of parameter variation $\alpha=10$ for the first two PR bands, $m=1$ (a) and $m=2$ (b). Each resonance tongue stems exactly from the detuning predicted by Eq.~\eqref{eq:NLSreson1} and the maximum gain (shown in the insets) only slightly drifts away from that value. Our analytical estimates refer to the first resonance band  and are shown in Figs.~\ref{fig:alpha10normal}(a) and \ref{fig:alpha10anomal}(a) to be  quite accurate. Instead in Figs.~\ref{fig:alpha10normal}(b) and \ref{fig:alpha10anomal}(b) only numerical results are provided. We observe that the second order PR exhibits a threshold value for $h$ below which the instability gain is virtually zero. This occurs also for higher-order PR and both in normal and anomalous GVD.
As explained above this is not due to the first derivative in Eq.~\eqref{eq:linODE2} which is not a damping term, but can be qualitatively ascribed to the the fact that in the Hill equation the forcing function contains several overtones of $\alpha$ and they could in special cases suppress completely a subset of resonances; see \cite{VerhulstEnc} and references therein.

\begin{figure}
	\centering
		\includegraphics[width=0.70\textwidth]{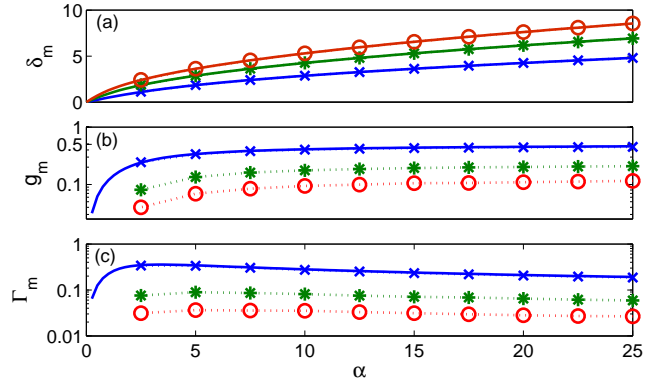}
	\caption{Characterization  of the first three PR peaks ($m=1,2,3$) as a function of $\alpha$.  Normal GVD ($s_{0}=+1$), $s_1 = -1$, $n_1 = 1$ and $h = 0.5$.  We plot in (a) the resonant frequency calculated in Eq.~\eqref{eq:NLSreson1} and obtained from the ODE solution as the point which maximizes the instability gain, which is shown, in logarithmic scale, in (b), which includes  also the solution of Eq.~\eqref{eq:gain1}, for $m=1$. Finally in (c) the instability bandwidth calculated numerically by means of the Hill determinant method and analytically for $m=1$, according to Eq.~\eqref{eq:bandwidth1}. In every panel the same convention is used, i.e.~solid lines represent analytical results, while dotted lines with markers are obtained by numerical  calculations: specifically the blue dotted line with crosses is associated to $m=1$, the green dotted line with stars to $m=2$ and the red dotted line with circles to $m=3$. The solid lines use the same color convention. }
	\label{fig:alphaiternormal}
\end{figure}

\begin{figure}
	\centering
		\includegraphics[width=0.70\textwidth]{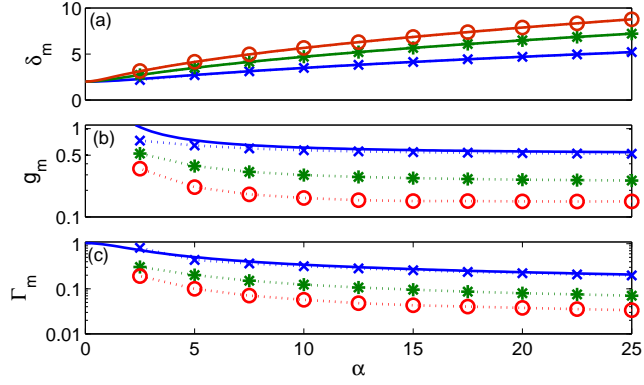}
	\caption{Same of Fig.~\ref{fig:alphaiternormal} in the case of anomalous GVD ($s_{0}=-1$) and $s_1=+1$.}
	\label{fig:alphaiteranomal}
\end{figure}

We provide also the curves of resonant frequency, gain and bandwidth as a function of $\alpha$ for fixed amplitude of the variation of nonlinearity and dispersion, $|s_1| = n_1 =1$ and $h=0.5$, Figs.~\ref{fig:alphaiternormal} and \ref{fig:alphaiteranomal}. The forcing terms $s_1$ and $n_1$ are of equal amplitude as above, and the perturbation is quite large. Nevertheless our estimates for the first band prove quite reliable; moreover the resonant frequency  hardly differs from its $h\to 0$ value at any resonant peak.

We notice that in the normal GVD regime (Fig.~\ref{fig:alphaiternormal}) each resonant frequency converges to zero as $\alpha\to 0$ while the gain $g\to h^m$ from below as $\alpha\to\infty$. Anomalous  dispersion (Fig.~\ref{fig:alphaiteranomal}) causes gain to approach the same limit but from above, while the resonant frequency is limited by the conventional MI band which cuts off at $\delta=2$.

\subsection{Including higher order effects}
\label{sec:HOD}

To conclude this section we briefly discuss how to compute the resonant frequency in a generalized NLS model including higher order terms. As above the linearized equation can be cast as
\begin{equation}
i\dot{\ket{\phi}} = H_{\rm gnls}(z) \ket{\phi},\;H_{\rm gnls} \equiv H^0+\tilde{H}(z).
\label{eq:GNLSlin}
\end{equation}
where we split as before the Hamiltonian in two parts, a constant and an oscillating one. We also assume that any common diagonal term of $H_{\rm gnls}$ is eliminated by a suitable change of variable, which is always possible.
At that point we have a traceless $2\times2$ matrix, thus if we define $\pm\omega_{0}$ to be the  eigenvalues of  $H^0$, and assume they are both real (for a certain choice of parameters and detuning), the resonance condition becomes $\omega_0=\sqrt{c_1^0c_2^0}=\alpha/2$ and from the expression of $\omega_0$ we derive the corresponding detuning.

We consider a generalized model which includes higher-order dispersion (HOD) terms up to the fourth order. It is well known that only even HOD terms contribute to MI \cite{AgrawalNL}. We  thus just need to modify  $c_{1,2}$ in Eq. \eqref{eq:NLS1PQ} by making the following substitution
\[
-\frac{1}{2}s(z)\delta^2 \mapsto -\frac{1}{2}s(z)\delta^2 -\frac{1}{24}\beta_4^n\delta^4,
\]
where $\beta_4^n$ is the normalized  fourth order dispersion (FOD) and is assumed to be constant. The third order dispersion (TOD) appears only as a common diagonal term and can be removed. The resonance condition is thus a quartic equation in $\delta^2$ and if $\beta_4^n<0$ there may  exist more than just one pair of sidebands which undergo parametric resonance. This is analogous to the conventional MI in the presence of HOD, see e.g.~\cite{Biancalana2003}.

%For example if we include self-steepening,  we notice that whatever small $\alpha$ there exist a sequence of resonances, in contrast to what appeared in Ref.~\cite[Fig.~1(b)]{Abdullaev1996}; the resonant detuning is indeed
%\begin{equation}
%	\delta = \frac{1}{\left|s_0\right|} \sqrt{-2n_0P_0(s_0+n_0P_0\tau^2)\pm\sqrt{\left[2n_0P_0(s_0+n_0P_0\tau^2)\right]^2+\left(m\alpha s_0\right)^2}}
%	\label{eq:GNLSreson2}
%\end{equation}
%where $\tau$ is the coefficient of the usual self-steepening term which here is supposed to be multiplied by the overall varying nonlinear coefficient.

\section{Comparison with numerical solution of NLS}

In order to assess the correctness of our analysis we solve Eq.~\eqref{NLS1} by means of the split-step Fourier method. We use the same parameters as above: $s_0=1$, $\alpha=10$, $-s_1=n_1=1$ and $h=0.5$. We use a large $\alpha$ in order to achieve widely separated PR resonances and large gain in the normal GVD regime, see Fig.~\ref{fig:alphaiternormal}. Finally we operate under normal GVD, because anomalous GVD gives rise to the conventional MI bands which have a larger gain (four times larger than the PR gain in the case of $h=0.5$).

\begin{figure}
	\centering
		\includegraphics[width=0.85\textwidth]{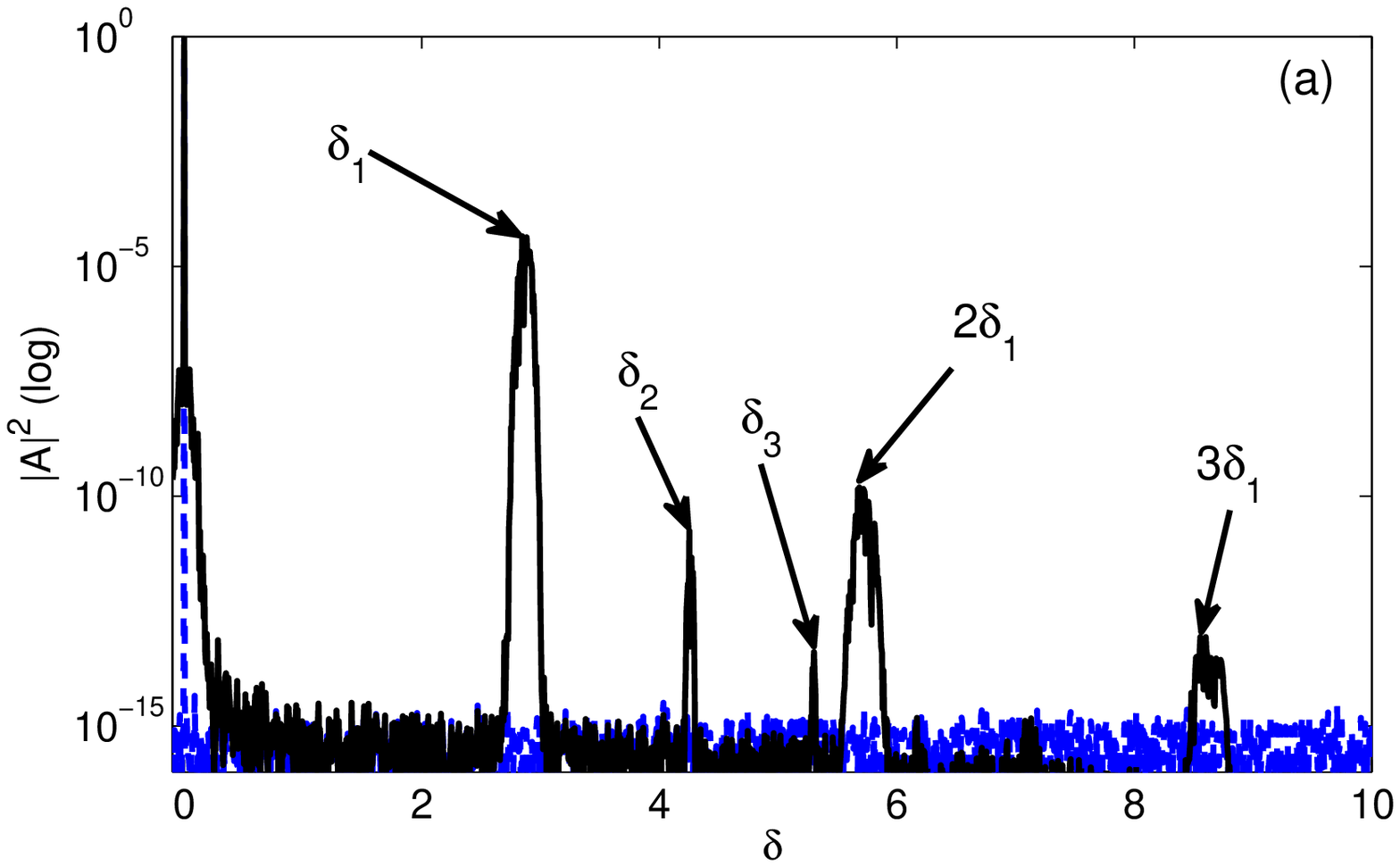}
		\includegraphics[width=0.85\textwidth]{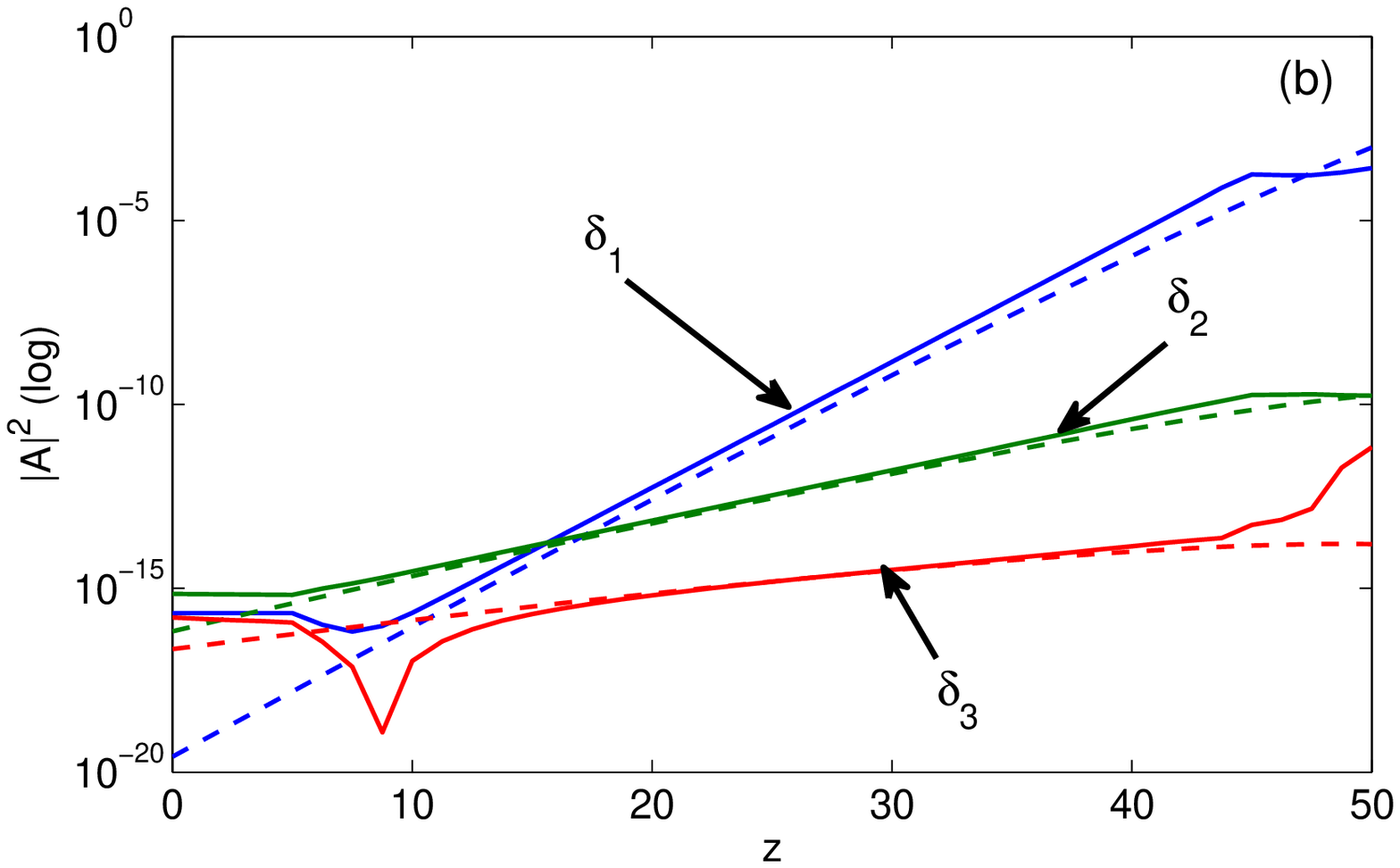}
		\includegraphics[width=0.85\textwidth]{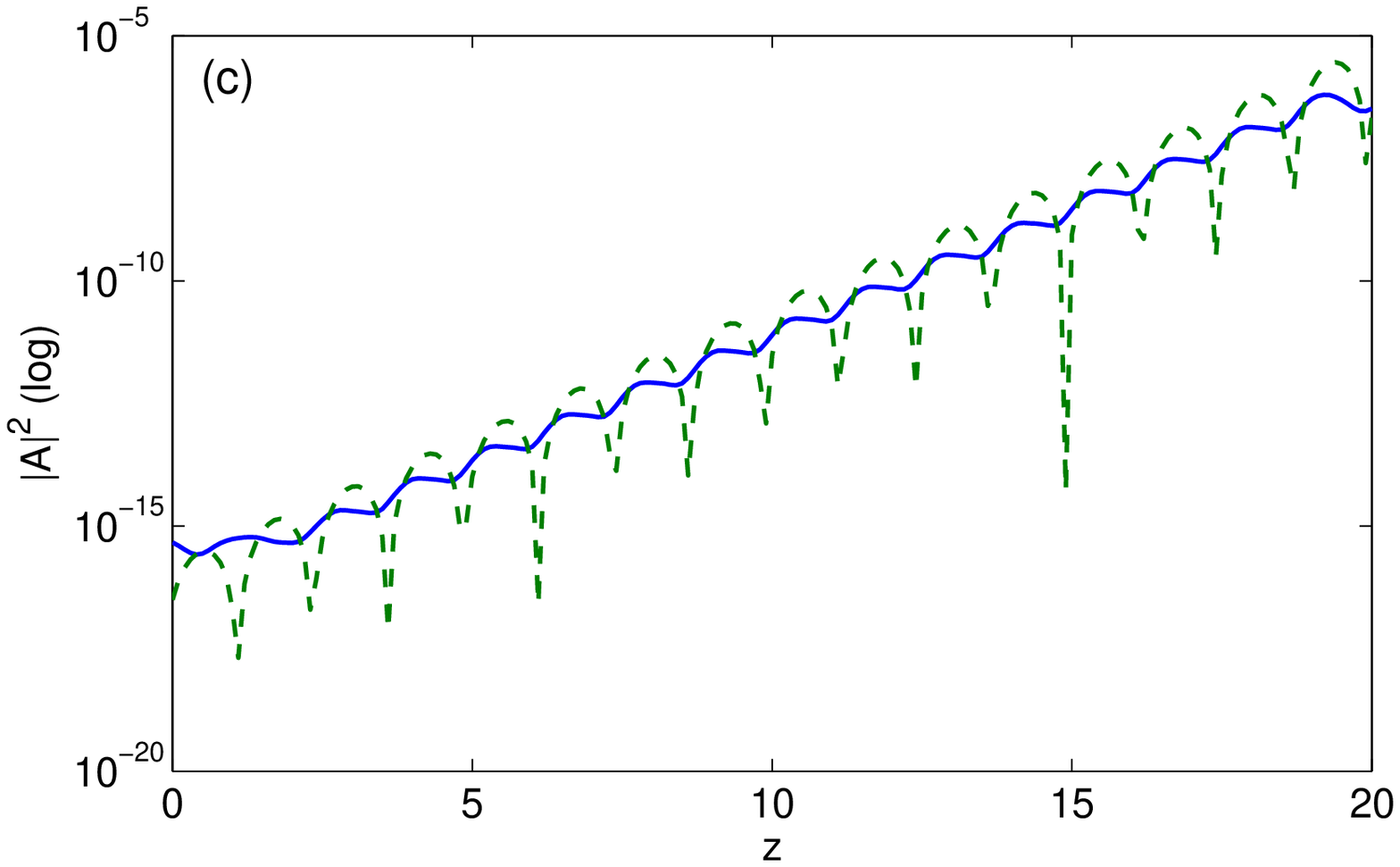}
	\caption{Numerical evolution of the NLS equation: (a) comparison of input (blue dashed line) and output (black solid) spectra. GVD is normal and the total propagation length is $z=38$, $\alpha=10$, $h=0.5$ and $n_1=-s_1=1$. The $m=1,\,2,\,3$ PR peaks as well as  the first two mixing products of $m=1$ are highlighted by arrows. (b) Amplification of the first three peaks: extracted from the spectrum (solid line) and predicted by the linearized analysis, i.e.~exponential growth with gain $g_m$ (dashed line). (c) The detail of the amplification process on a shorter scale: we use a different set of parameters, $\alpha=5$ and $h=0.9$ (with $n_1=-s_1=1$ as above) in order to have higher gain and a larger period. 
Blue solid line represents the evolution of the 1st PR peak spectral component, the dashed green line  the solution of the averaged equations, Eq.~\eqref{eq:avansatz}. The amplification-deamplification process is apparent and agrees with the prediction of the averaged linear equation. }
	\label{fig:NLSFFT}
\end{figure}

In Fig.~\ref{fig:NLSFFT}(a) we show the output spectrum after a propagation distance $z=38$. One can clearly distinguish the first three peaks ($m=1,\,2,\,3$), plus two peaks corresponding to the four-wave mixing of the carrier and the $m=1$ PR band. In table \ref{tab:alpha10} we report the values of peak detuning and bandwidth extracted from the spectrum of Fig.~\ref{fig:NLSFFT}(a), the numerical results of the linearized problem discussed in the previous section and predictions obtained by phase-matching considerations, Eq.~\eqref{eq:grating}. The peak positions are in very good agreement with the results of the previous section. The approximate phase-matching always underestimates the value of the actual detuning. 

%. Moreover, in order to attain to the large values of resonant detuning we present in the next section, a large $\alpha$  is required. This unfortunately does not allow to fully appreciate the improved accuracy of our estimates. {\bf This last sentence must be removed or rewritten}

% From table \ref{tab:alpha10} we also notice that the predicted bandwidth is  in good agreement with our numerical simulations. The first PR band exhibits a small displacement of the peak gain and a larger bandwidth. The explanation is that it has a wider bandwidth with respect to the higher order PR peaks, thus it is easier to notice the influence of noise. {\bf what does it mean?}

In Fig.~\ref{fig:NLSFFT}(b) we show how the three PR peaks evolve during propagation: the instability leads to amplification at the rate predicted by the linearized problem, i.e.~Eq.~\eqref{eq:gain1} (see the dashed lines), apart from  saturation occurring towards the end of the evolution.

Next we discuss the small scale oscillations reported in \cite{Mussot2012,Mussot2012b}, i.e.~an  amplification-deamplification cycle undergone by the resonant peaks. 
Since Fig.~\ref{fig:NLSFFT}(b) does not allow us to visualize them, it is useful to introduce a new set of parameters which gives larger gain, so that fewer periods are sufficient. Thus we choose $h=0.9$ and normal GVD conditions. Moreover, in order to accurately visualize the oscillations, we set a larger period, i.e.~$\alpha=5$. 
In Fig.~\ref{fig:NLSFFT}(c) we show the evolution of the first peak, together with the solution of the averaged problem, Eq.~\eqref{eq:avansatz}, with $\omega_0=\alpha/2$ (a phase shift is applied so that the two curves overlap). The numerical and approximate solutions of the linear problem agree quite well, thus providing a good explanation of the process without explicitely resorting to a three wave dynamical system (as in \cite{Mussot2012b}). The amplification-deamplification depends in practice exclusively on $\alpha$, i.e.~it stems from the forced oscillations impressed on resonance by the variation of parameters, superimposed to the unstable growth. This is completely analogous to a pendulum the pivot of which is displaced periodically: a small displacement from the position in which the bob points down initiates oscillations (at the natural frequency) which are further amplified at each cycle.

For the sake of completeness we compare, in table \ref{tab:alpha5}, the position of resonant peaks with the predictions of the grating phase matching and the Hill equation. Our estimate Eq.~\eqref{eq:NLSreson1}  is the closest to numerical NLS result despite it is of zeroth order in $h$, while the simple phase-matching argument, Eq.~\eqref{eq:grating}, is evidently less and less accurate as $\alpha\to0$.

%%%%%%%%%%%%%%%

% tables
\begin{table}
	\centering
		\begin{tabular}{|c|ccc||cc|}
		\hline
		$m$&\multicolumn{3}{|c||}{	$\delta_m$} &\multicolumn{2}{|c|}{	$\Gamma_m$}  \\
		\hline
		& NLS & Hill \eqref{eq:NLSreson1}& Grating \eqref{eq:grating} & NLS & Hill (num.)\\
		\hline
			1& 2.8463 & 2.8632 & 2.8284 & 0.34558 & 0.27481\\
			 \hline
			2&  4.2537 & 4.2544 & 4.2426 & 0.087965 & 0.081239\\
			  \hline
			3&  5.303 & 5.2978 & 5.2915 & 0.050265 & 0.035274 \\
			  \hline
		\end{tabular}
		\caption{Values of resonant frequencies and bandwidth for $\alpha = 10$. Comparison between data extracted from NLS evolution and the results of linearized model (solved both as phase matching condition and Hill equation). The resonant frequency is estimated at any order analytically, while numerical values are reported for bandwidth. }
	\label{tab:alpha10}
\end{table}
\begin{table}
	\centering
		\begin{tabular}{|c|ccc|}
		\hline
		$m$&\multicolumn{3}{|c|}{	$\delta_m$} \\
		\hline
		& NLS & Hill \eqref{eq:NLSreson1}& Grating \eqref{eq:grating} \\
		\hline
			1&     1.8912  &  1.8399  &  1.7321\\
			 \hline
			2& 2.9719   & 2.8632  &  2.8284\\
			  \hline
			3&  3.6191 &   3.6239  &  3.6056 \\
			  \hline
		\end{tabular}
		\caption{Same as in Tab.~\ref{tab:alpha10}, but for $\alpha = 5$.}
	\label{tab:alpha5}
\end{table}

%%%%%%%%%%%%%%%%%%%%%%%%%%%%%%%%%%%%%%%%%%%%%%%%%%%%%%%%%%%%%%%%
% PCF design and simulations
%%%%%%%%%%%%%%%%%%%%%%%%%%%%%%%%%%%%%%%%%%%%%%%%%%%%%%%%%%%%%%%%%%
\section{The design of a periodically tapered PCF}

In this last section we propose a feasible system which supports PR instability peaks exhibiting relatively large frequency detunings. This is important, e.g., in quantum optics where the implementation of new sources of entangled photons with a reduced Raman decorrelation is of great practical interest, and this can provide an alternative approach to what proposed in Refs.~\cite{Rarity2005,Silberhorn2010}.

In our calculations we use a periodically tapered PCF whose index contrast, small core and design flexibility allow to obtain a large nonlinearity, together with regions of small GVD. A similar system was already used in \cite{Mussot2012,Mussot2012b}, but here we predict the possibility of observing far detuned, tunable instability peaks, by employing short tapering periods.

At first we explore the design space (pitch and filling fraction) in order to operate in a region of small normal GVD near  $\lambda_0=1064$ nm, and such that the zero-dispersion point (ZDP) can be approached by slightly varying the PCF geometry. The modal analysis is performed by means of COMSOL Multiphysics \cite{comsol}.

\begin{figure}
	\centering
		\includegraphics[width=0.70\textwidth]{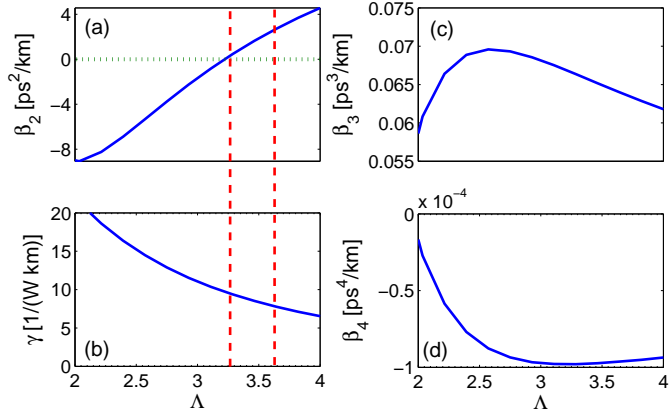}
	\caption{Main parameters of a PCF as a function of pitch $\Lambda$ at wavelength $\lambda_0 = 1064$ nm. The fiber is made of pure silica and the filling fraction $d/\Lambda=0.4$. The red dashed lines in (a,b) show the range spanned by the parameters when tapering the fiber. }
	\label{fig:PCF}
\end{figure}

In Fig.~\ref{fig:PCF} we show the properties of a PCF made of pure silica with a triangular lattice of air holes. We assume the air filling fraction $d/\Lambda=0.4$ ($d$ is the air hole diameter, $\Lambda$ is the pitch) to be constant so that by varying $\Lambda$,  $d$ is adjusted accordingly. We are interested mainly in two quantities: the GVD ($\beta_2$), see Fig.~\ref{fig:PCF}(a), and the nonlinear coefficient ($\gamma$), Fig.~\ref{fig:PCF}(b), calculated according to the full vectorial model of the effective area, see \cite{Afshar09}. Fig.~\ref{fig:PCF} also shows the next two terms of the Taylor expansion of the GVD, $\beta_{3}$ [Fig.~\ref{fig:PCF}(c)] and $\beta_4$ [Fig.~\ref{fig:PCF}(d)], as functions of the pitch $\Lambda$.

We propose to operate between $\Lambda_{\rm min} = 3.25$ $\mu$m and $\Lambda_{\rm max} = 3.6$ $\mu$m. In this range, $\beta_2\in [0.4,2.7]$ ps$^2/$km and $\gamma\in [7.8,9.4]$ W$^{-1}$/km. 
%At $\Lambda_{\rm min}$ our wavelength of operation is close to the ZDP, while at $\Lambda_{\rm max}$. 
Thus the GVD undergoes, when compared to its average value, very large oscillations, equivalent to $h\approx0.85$ in Eq.~\eqref{eq:dispnlcos}. Note that we do not cross into the negative GVD region in order to inhibit any spurious occurrence of the classical MI.  Moreover the dependence of GVD on $\Lambda$ is, in good approximation, linear. Thus a  cosine-shaped tapering of the PCF leads to a cosine variation of GVD. The value of $\gamma$ is only slightly modified by the tapering and can be also approximated as a cosine. This allows to straightforwardly apply the theory developed in the previous sections. We thus use the following simple formulas for the variations of the parameters:
\[
\Lambda = \Lambda^0 + \Lambda^1 \cos{\tilde\alpha z},\;
\beta_2 = \beta_2^0 + \beta_2^1 \cos{\tilde\alpha z},\;
\gamma = \gamma^0 + \gamma^1 \cos{\tilde\alpha z},\;
\]
where all the parameters are reported in table \ref{tab:PCF}. 
We use as above the superscripts $0$ and $1$ to denote the average value and first Fourier coefficient, and we use $\tilde\alpha$ for the dimensional spatial frequency of tapering ($\tilde\alpha =\alpha/Z_{NL}$).
As far as HOD terms are concerned, they are in a good approximation constant and are also reported in table \ref{tab:PCF}. Finally in our simulations we include the self-steepening term and stimulated Raman scattering response of silica, see \cite{AgrawalNL}, in order to obtain a very realistic simulation.

\begin{table}
	\centering
		\begin{tabular}{|c|r|}
		\hline
		$d/\Lambda$ & 0.4\\
		$P_t$ 			& 25 W\\
		$\Lambda^0$ & $3.425$ $\mu$m\\
		$\Lambda^1$ & $0.175$ $\mu$m\\
		$\beta_2^0$ & $1.3695$ $\mathrm{ps^2/km}$\\
		$\beta_2^1$ & $1.1275$ $\mathrm{ps^2/km}$\\
		$\gamma^0$ & $8.7302$ /(W km)\\
		$\gamma^1$ & $-0.8140$ /(W km)\\
		$\beta_3$	& $6.43\times 10^{-2}$ $\mathrm{ps^3/km}$\\
		$\beta_4$	& $-0.97\times 10^{-4}$ $\mathrm{ps^4/km}$\\
		$\tilde\alpha$ 	& $47.64$ $\mathrm{m^{-1}}$\\		
		$L_{\rm taper}$ & $225$ m\\
		$L_{\rm tot}$ & $250$ m \\
		number of periods & $1708$\\ 
		\hline	
		\end{tabular}
		\caption{PCF parameters used in the generalized NLS model.}
	\label{tab:PCF}
\end{table}

We set our design target: the period of variation along the direction of propagation ($T_Z \equiv 2\pi/\tilde\alpha$) has to be such that the $m=1$ PR band is located at $\Delta f = \pm35$ THz. This quite large detuning allows to clearly distinguish PR from Raman gain, in contrast with \cite{Mussot2012} where PR peaks occur in the Raman gain spectral region and thus are further amplified by it (in the cited paper the spectrum exhibits a frequency asymmetry which is a clue to the enhancement of sidebands provided by Raman amplification).

We verified that self-steepening and Raman effects are of little importance in our case, nevertheless at large detuning we cannot neglect the effect of HOD. As discussed above, in sec.~\ref{sec:HOD}, only even order terms contribute to the instability, and amount to a simple modification of $c_{1,2}$ in Eq.~\eqref{eq:NLS1PQ}. 
In order to complete our design, we thus set $P_t = 25$ W, correct the PR condition including $\beta_4$ and finally obtain $\tilde\alpha = 47.7$ m$^{-1}$, corresponding to a period of tapering of $T_Z = 13.2$ cm. Neglecting $\beta_4$, the estimated period would be less than $10$ cm. The predicted gain at $\Delta f = \pm35$ THz is approximately $g\approx100$ km$^{-1}$. We simulate a $250$ m long fiber with a periodically tapered central part of about $225$ m, which corresponds to $1708$ periods. The power level and fiber tapering periods are only slightly more demanding than those used in \cite{Mussot2012}. The use of a highly nonlinear fiber (for example by using materials other than fused silica) can scale down power levels and thus nonlinear lengths conveniently, but this is beyond the scope of the present work.

\begin{figure}
	\centering
		\includegraphics[width=0.70\textwidth]{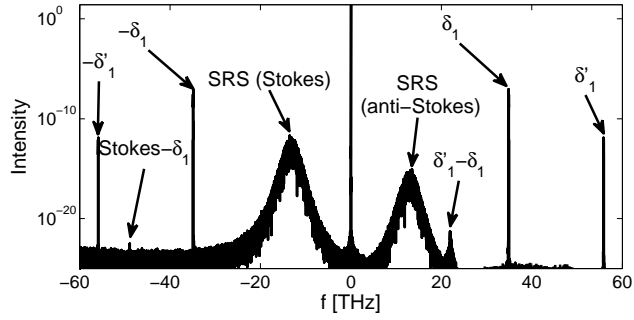}
	\caption{Output spectrum after the propagation in a periodically tapered PCF. All fiber parameters are listed in Tab. \ref{tab:PCF}. We detect the two Raman gain bands at $\Delta f = \pm13$ THz, Stokes (red-detuned) and anti-Stokes (blue-detuned), the former exhibits as expected a larger gain. We also label the main two PR instability peak pairs which correspond respectively to the design requirement of $\Delta f = 35$ THz and  to the additional phase matching allowed for by FOD.}
	\label{fig:GNLS}
\end{figure}

The spectrum at the end of the propagation is shown in figure \ref{fig:GNLS}. We notice two main peak pairs appearing beyond Raman-gain bands (the two broad asymmetric peaks at $\Delta f = \pm13$ THz). The first pair is located at $\Delta f = \pm34.8$ THz, the second pair at $\Delta f = \pm56$ THz. The bandwidth of each individual peak belonging to the first pair is $0.25$ THz, and the gain agrees well with our theoretical prediction ($g\approx100$ km$^{-1}$). The second PR peak pair exhibits smaller gain ($g'\approx70$ km$^{-1}$) and narrower bandwidth, and can be ascribed to FOD. which allows for an additional solution of the nonlinear phase matching condition, see e.g. \cite{Biancalana2003}. In summary, the presence of $\beta_4<0$ has the advantage of leading to a  period $T_Z$ larger than that predicted neglecting all HOD terms. It has also a drawback  that an additional PR band appears, but we verified this applies, for our specific choice of parameters, only at the first PR order. If we include losses in our simulations we obtain a smaller available power, which affects mainly the peak gain, see Eq. \eqref{eq:gain1}. Indeed  the resonant detuning, see Eq. \eqref{eq:NLSreson1}, is nearly independent on input power for $\alpha\approx200$, which corresponds to our $\tilde\alpha$. 

%To summarize, the presence of $\beta_4<0$ has two effects: a positive and a negative one {\bf positive and negative should be replaced by advantage/disadvantage or drawback}. The period can be larger than those predicted in the previous section neglecting such a term, which is good {\bf Terms like good, bad or similar must not appear in a scientific paper. I suggest rewriting completely this paragraph, the point of which is not understandable at all.}. By contrast an additional band pair appears  but this occurs, for our specific choice of parameters, only at the first PR order. At such a large $\tilde\alpha$, including linear losses in the simulation does not affect much the resonant frequencies (see Eq. \eqref{eq:NLSreson1} with $\alpha\approx200$), but strongly reduces the gain, due to the reduced available power, Eq.~\eqref{eq:gain1}. The structure of the spectrum at larger detuning is quite complicated by the occurrence of higher order PR and the four-wave mixing products of  unstable frequencies with pump and among themselves.

\section{Conclusion}

In this work we have presented a thorough analysis of parametric resonance instabilities occurring in a generalized nonlinear Schr{\"o}dinger equations with varying dispersion and nonlinearity. We have shown that, in the  case of GVD and nonlinear coefficients varying in a simple sinusoidal way, it is possible to predict the maximum gain and instability margins. The calculation of the resonant frequencies is possible even in more general cases, since it depends only on the average values of coefficients and the period of variation. Our calculations provide analytical estimates and are more accurate than those found in previous literature. We have validated our theory by means of numerical dynamical simulations and found a very good agreement. Finally we have designed a periodically tapered photonic crystal fiber which allows to achieve instability peaks at a large tunable frequency detuning from a given pump wavelength. Several higher-order resonant peaks are present but their gain and bandwidth are generally smaller than the one occurring at the smallest detuning. Such a system can be useful in quantum optical applications such as the efficient generation of entangled photon pairs in regions of frequencies far from the Raman gain peak.

\section*{Acknowledgments}
This research is funded by the German Max Planck Society for the Advancement of Science. F.~B.~would like to thank Arnaud Mussot for useful discussions.

\end{document}